\documentclass[twoside,11pt,letterpaper]{article}
\usepackage[utf8]{inputenc}

\usepackage[letterpaper,top=2.54cm,bottom=2.54cm,left=2.54cm,right=2.54cm,marginparwidth=1.75cm]{geometry}
\usepackage{amsmath}
\usepackage{graphicx}
\usepackage{epstopdf, epsfig}
\usepackage{svg}
\usepackage{natbib}
\usepackage{authblk}   
\usepackage[colorlinks,citecolor=blue,urlcolor=blue,linkcolor=blue,bookmarks=false]{hyperref} 
\usepackage{booktabs}

\title{Weierstrass' variational theory for analysing meniscus stability in ribbon growth processes}

\author[1]{Eyan P. Noronha\footnote{Corresponding author: \texttt{enoronha@andrew.cmu.edu}}}
\author[1]{German A. Oliveros}
\author[1]{B. Erik Ydstis}
\affil[1]{Department of Chemical Engineering}
\affil[ ]{Carnegie Mellon University}
\date{November 7, 2020}

\begin{document}

\maketitle

\begin{abstract}
We use the method of free energy minimization to analyse static meniscus shapes for crystal ribbon growth systems. To account for the possibility of multivalued curves as solutions to the minimization problem, we choose a parametric representation of the meniscus geometry. Using Weierstrass' form of the Euler-Lagrange equation we derive analytical solutions that provide explicit knowledge on the behavior of the meniscus shapes. Young's contact angle and Gibbs pinning conditions are analysed and shown to be a consequence of the energy minimization problem with variable end-points. For a given ribbon growth configuration, we find that there can exist multiple static menisci that satisfy the boundary conditions. The stability of these solutions is analysed using second order variations and are found to exhibit saddle node bifurcations. We show that the arc length is a natural representation of a meniscus geometry and provides the complete solution space, not accessible through the classical variational formulation. We provide a range of operating conditions for hydro-statically feasible menisci and illustrate the transition from a stable to spill-over configuration using a simple proof of concept experiment. 
\end{abstract}


\section{Introduction}
\textit{These investigations, which have found their confirmation in striking agreement with careful experiments, are among the most beautiful enrichments of natural science that we owe to the great mathematician.}

--- Carl Fredrich Gauss on Laplace's theory of capillary action, which was later refined by him into its modern variational form.
\vspace{6pt}

\noindent The Young-Laplace equation was developed by \citet{young1805iii}, who provided a qualitative theory for surface tension, and \citet{laplace1799traite}, who mathematically formalized the relationship described by Young. This theory was later refined by \citet{gauss1877principia} using Bernoulli's principle of virtual work. Using the fundamental principles of dynamics, he derived the Young-Laplace equation and Young's contact angle condition from a single variational framework. He argued that the energy of a mechanical system in equilibrium is unvaried under arbitrary virtual displacements consistent with the constraints. This spirit of variational analysis is still used in practice to describe the meniscus shape in interface problems.

The existence of a static meniscus plays a critical role in capillary-shaped ribbon growth systems such as the Dendritic web growth \citep{seidensticker1980silicon}, Edge-defined Film Growth \citep{rossolenko2019numerical}, Low Angle Silicon Sheet  growth~\citep{jewett1982low}, and the Horizontal Ribbon Growth process~\citep{Oliveros2015}. In these processes, the ribbon shape is supported by the formation of a stable meniscus and grows without touching any external surface like the crucible wall. This prevents the formation of defects during solidification and allows for the formation of single crystal ribbons \citep{duffar2010crystal}. 

\begin{figure}
  \centerline{\includegraphics[width=0.65\linewidth]{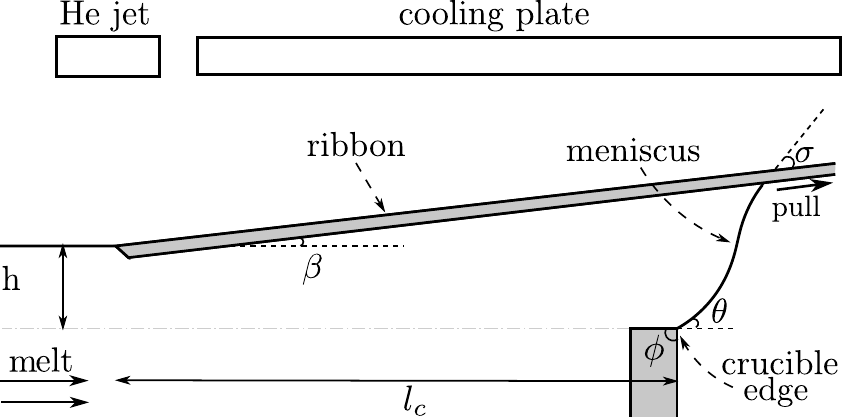}}
  \caption{Schematic for a horizontal crystal ribbon growth process. The formation of a meniscus at the end of the crucible is essential for steady state operation}
  \label{fig:HRG_meniscus}
\end{figure}

Figure~\ref{fig:HRG_meniscus} describes the schematic of a Horizontal Ribbon Growth (HRG) process which will serve as an example to illustrate the application of our theory. A bath of molten substrate is cooled from the top to form a thin ribbon of single crystal which is continuously extracted. A narrow Helium cooling jet is used to provide intense cooling for solidification and keeps the starting point of the ribbon almost fixed for all feasible pull speeds~\citep{Helenbrook2016}. A seeding process takes place at the outlet while the melt is being continuously replenished at the other end. Thin sheets of single crystal can be pulled at relatively high speeds due to enhanced heat transfer with the surroundings~\citep{Thomas1987}. This provides an advantage over the present crystal growing methods, like the Czochralski process, where the sheets are prepared by slicing a single crystal boule followed by tedious, time consuming grinding and lapping operations which result in a large percentage of the original crystal being wasted~\citep{william1962process}. The weight of the ribbon is supported by the melt, which forms a meniscus between the ribbon and the edge of the crucible, thereby reducing the mechanical stresses on the crystal. 

\begin{figure}
  \centerline{\includegraphics[width=.7\linewidth]{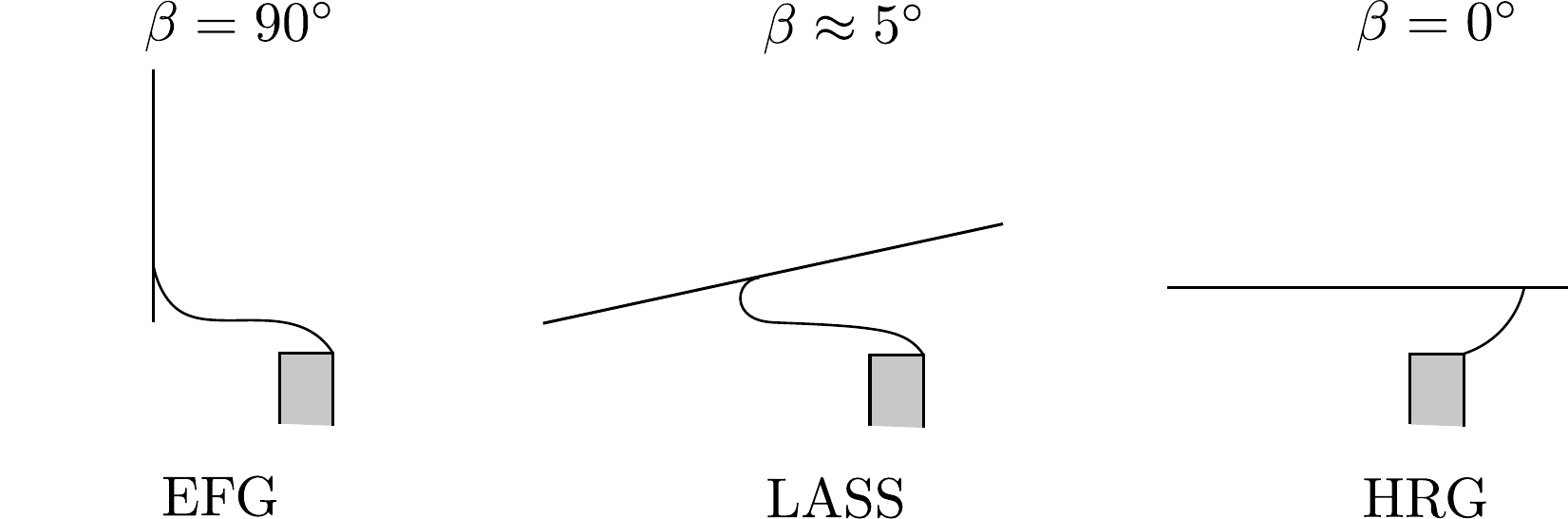}}
  \caption{Characterization of the ribbon growth family based on pull angle ($\beta$)}
  \label{fig:ribbon_family}
\end{figure}

Several well known ribbon growth techniques can be characterized by the angle ($\beta$) at which the ribbon is pulled from the melt as illustrated in Figure~\ref{fig:ribbon_family}. For the case when $\beta=90^\circ$, the ribbon is pulled perpendicular to the surface of the melt and relates to the family of vertical ribbon growth techniques such as the Edge-defined Film Growth (EFG), and the Dendritic web (WEB) growth processes. In the EFG process, the role of the crucible in the previous example is substituted by a melt-wettable die which provides a pinning boundary for the meniscus. The die determines the shape of the meniscus and thus the cross section of the growing crystal ribbon~\citep{yang2006meniscus}. This makes it important to understand the meniscus geometry in order to study its effect on crystal shape and quality.

Another family of ribbon growth methods, characterized by their low pull angles, are the Low Angle Silicon Sheet (LASS) process, and the Horizontal Ribbon Growth (HRG) process~\citep{bates1984low,kellerman2013floating}. These methods have the advantage of having a large solid-liquid interfacial area making it easier for the latent heat to dissipate and leading to higher production speeds. However, two technical issues appear while operating these processes: the ribbon freezing onto the crucible (down-growth) and the melt spilling-over the crucible (spill-over)~\citep{Kudo1980,bleil1969new,kellerman2010floating}. These two issues are directly related to the formation of a short or unstable meniscus between the ribbon and the crucible edge~\citep{Daggolu2012,Daggolu2013,Oliveros2015}.

The main objective of this paper is to analyse these instabilities of the meniscus for the horizontal ribbon growth processes, while at the same time keeping the analysis general enough to be extended to other problems of physical or engineering importance.

The meniscus profiles for HRG were first investigated by \citet{rhodes1980investigation} around the same time as \citet{Kudo1980} performed his first HRG experiments with silicon. They developed a mathematical model based on hydrostatics, to describe the shape of the meniscus that must be formed between the ribbon and the crucible edge. They found that the hydrostatically feasible configurations require the meniscus to be ``taller" than the melt height, and that the ribbon be pulled at a slight angle, which coincided with Kudo's experimental operation. In 2012, \citet{Daggolu2012,Daggolu2013,daggolu2014analysis} constructed a thermal-capillary model describing the interaction between fluid flow and heat transfer in a HRG system. In their results they captured the critical nonlinearities in the system, such as the existence of multiple steady states for a given pulling speed. The problems of melt spilling over and freezing to the crucible were assessed by doing a sensitivity analysis on the length of the meniscus as a function of melt height and pull angle~\citep{Daggolu2013}. Multiplicity of steady states with respect to pull speeds have also been observed in the case of EFG process \citep{yeckel2013analysis, samanta2011analysis}. This multiplicity is manifested as two states of the ribbon thickness for the same pull speed but with different failure limits~\citep{yeckel2016modeling}.

Classical variational analysis has proven useful in the study of meniscus stability for capillary based processes. For the Czochralski process, \citet{mika1975shape} used free energy minimization along with the concepts of variational calculus to numerically determine instability conditions for the meniscus. \citet{mazuruk2013static,mazuruk2016dynamic} addressed the static stability problem for the Bridgman process using numerical simulations to calculate the sign of the second order variations for the governing free energy formulation. Using the observation that the angle between the crystal and the melt should converge towards a constant value, $11^\circ$ for the case of silicon \citet{surek1976theory} developed a theory for shape stability in capillary shaped crystal growth systems based on deviations from this angle. Dynamic and static stability was numerically addressed by~\citet{tatartchenko2010possibility} using Lyapunov based techniques and variational principles.

Outside the area of crystal growth, variational principles have been used to prove the existence and stability of menisci shapes for different capillary geometries~\citep{wang2012capillary}. A similar approach has been used by \citet{pitts1973stability} and \citet{vogel1987stability} to study the shape of liquid pendant drops and identify regions of stability before the drop breaks. \citet{soligno2014equilibrium} implemented a numerical method to minimize thermodynamic potential function and calculate the interface shape of liquids for various wall geometries. \citet{lawal1982stability,lawal1982stability2} used polar coordinates in their variational formulation to obtain multiple critical solutions for their drop geometries on an inclined surface. They observed that for a fixed Bond number, axisymmetric sessile shapes on horizontal surfaces lose stability at a drop volume that corresponds to a point of bifurcation into a family of asymmetric shapes. As we shall see in Section~\ref{results}, the ribbon growth configuration also admits a point of bifurcation into stable and unstable families of menisci. 

Very recently \citet{Oliveros2015} used the classical variational approach to find existence and stability conditions for menisci in a HRG process. The analysis showed that stationary menisci arising as solutions to the classical Euler-Lagrange equation were stable as long as the solution satisfied the existence conditions. Due to the well known complexity and non-linearity of the ribbon growth systems, we were left with the question of whether or not the system had any unstable configurations that can't be captured by the traditional variational tools. In Weierstrass' variational theory this limitation is overcome by formulating the geometry of the meniscus in parametric form~\citep{bolza2018lectures}. This approach allows for the possibility to find stationary curves described by multi-valued functions (Figure \ref{fig:formulation}), which help expand the solution space. We refer the readers to \citet{patinomodeling} for further details on the development of the theory.

It is important to note that heat transfer and fluid flow also play a critical role in capillary shaped crystal growth processes~\citep{stelian2017numerical,le20193d}. However here we decouple these phenomena and focus only on the static stability of the meniscus in absence of heat transfer. In doing so, we are able to provide explanations to phenomena like multiplicity of steady states as observed in the EFG and HRG processes~\citep{Daggolu2013,yeckel2013analysis} and the existence of destabilizing multi-valued menisci in crystal growth systems~\citep{miroshnichenko2016effect,schwabe2011analysis}, among others, using a first principles approach. 

\section{Problem Statement}
\begin{figure}
  \centerline{\includegraphics[width=0.85\textwidth]{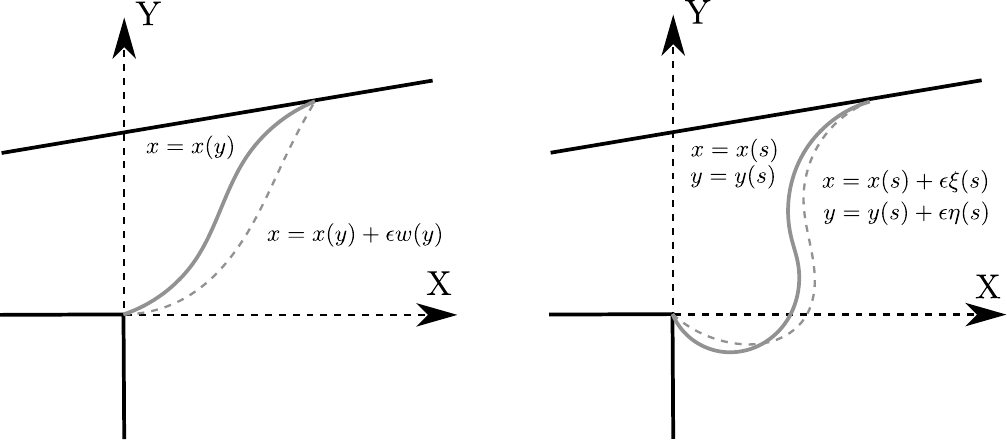}}
  \caption{(left) Stationary curves arising from the solution of the classical Euler-Lagrange equation yield single valued functions of the type $x(y)$. (right) Weierstrass' variational approach expand the solution space to account for multi-valued stationary curves, described by curves of the type $x(s)$ and $y(s)$.}
  \label{fig:formulation}
\end{figure}

The free energy ($\Delta U$) of the three phase system consisting of a rigid fixed ribbon, a gas and a liquid bounded by the interface $x=x(y)$, as seen in Figure \ref{fig:formulation}, is given by the expression~\citep{Oliveros2015}:
\begin{align}\label{eq:classical_energy}
    \Delta U=\int_0^H-\Delta P\: xdy+\gamma \sqrt{1+x'}dy=\int_0^HF(x,y,x')dy.
\end{align}
The right-hand side is divided into two terms. The first term represents the potential energy due to hydro-static pressure ($\Delta P$). The second term represents the free surface energy of the interface due to surface tension ($\gamma$). $H$ is the maximum height of the meniscus and corresponds to the length of integration in the vertical direction in the Cartesian co-ordinate system. The Euler-Lagrange equation for the variational problem that results from minimizing the free energy gives the Young-Laplace equation:
\begin{align}
    \Delta P+\gamma \frac{x''}{(1+x'^2)^{3/2}}=0.
    \label{Y-L eqn}
\end{align}
Analytic solutions have been developed for the Young-Laplace equation \eqref{Y-L eqn} in 2 and 3 dimensions using Legendre elliptic functions~\citep{Oliveros2015,alimov2016piercing}. However, in some cases the solution to geometric problems that use the classical variational formulation cannot be described by functions of the form $x = x(y)$ in a Cartesian coordinate system. For example, the multi-valued meniscus of a non-wetting sessile drop on an incline plane cannot be described using single-valued functions~\citep{bonartinfluence,lawal1982stability2}. A similar problem exists in describing the full spectrum of minimum energy curves for ribbon growth systems. The stationary curves arising from free energy minimization are often multi-valued and therefore require a less restricting and more ``natural" representation. By choosing the arc length as a parametric variable, we see the emergence of a natural representation of the meniscus shape which allows us to find the complete solution space. These interface curves are also shown to share similarities with the family of Euler's elastic curves.
\section{Free energy reformulation in parametric form} \label{reformulation}

Instead of denoting the interface in a Cartesian frame as $y=y(x)$ or $x=x(y)$, we represent them in parametric form $x(s)$, $y(s)$. The parameter $s$ is chosen such that $x(s)$ and $y(s)$ come out as single valued functions with respect to the parameter. The reformulated free energy expression with respect to the parameter $s$ is
\begin{align}
    \Delta U=\int_0^{s_t}-\Delta P \:xy'ds+\gamma \sqrt{x'+y'}ds=\int_0^{s_t}g(x,y,x',y')ds,
    \label{eq:freeenergy}
\end{align}
where $s_t$ is the total length of integration. The first term accounts for the hydro-static energy of the system. Due to the presence of gravity, the pressure in the liquid is given as a function of height, $\Delta P=\rho g_r(h-y)$, where $\rho$ is the density of the liquid and $g_r$ is the acceleration due to gravity. The second term represents the surface energy of the interface due to interfacial tension $\gamma$.

In performing this transformation, the value of $\Delta U$ must remain invariant for any type of parametric form chosen for $x$ and $y$.  Weierstrass showed that the necessary and sufficient condition for the invariance of $\Delta U$ is that the function $g$ be homogeneous and of degree one in the variables $x'$ and $y'$~\cite[p.~118]{bolza2018lectures}, i.e.
\begin{equation}
g(x,y,kx',ky') = k g(x,y,x',y'),
\label{eq:homogeneouscondition}
\end{equation}
where the prime represents differentiation with respect to $s$. From this homogeneity condition, there follow several relationships between the partial derivatives of $G$, which are useful in constructing the expressions for the first and second variation of $\Delta U$. 

Non-dimensionalizing length scales with respect to the capillary length, $\lambda_c=\sqrt{\gamma/(\rho g_r)}$ yields,

\begin{align*}
    \mathcal{U}(X,Y)=&\int_0^{S_t}(Y-H)XY'dS+\sqrt{X'+Y'}dS\nonumber\\ 
    =&\int_0^{S_t}G(X,Y,X',Y')dS,
\end{align*}
where
$\Delta U=\gamma \lambda_c\,\mathcal{U}, x=\lambda_c X, y=\lambda_cY, s_t=\lambda_cS_t \text{ and } h=\lambda_c H$.
The homogeneity condition is not affected by non-dimensionalization. 

\section{Stationary curves via the first variation} \label{first_variation}

The objective is to find conditions on the stationary curves $X(S), Y(S)$ that set the first variation of $\mathcal{U}$ to zero. This exercise leads to the well known Young-Laplace equation in parametric form. However, we briefly provide some important steps in the derivation as they will be useful in developing the forthcoming stability analysis.

Let $\epsilon\,\xi(S)$ and $\epsilon\,\eta(S)$ be small perturbations to the curves $X(S)$ and $Y(S)$, with $\epsilon$ as small as desired. The end points on the curve are kept fixed i.e. $\xi$ and $\eta$ are taken to be zero at the end points. The energy of this neighbouring curve is given by
\begin{align*}
    \mathcal{U}(X+\epsilon\,\xi,Y+\epsilon\,\eta)=\int_0^{S_t}G(X+\epsilon\,\xi,Y+\epsilon\,\eta,X'+\epsilon\,\xi',Y'+\epsilon\,\eta')dS.
\end{align*}
Applying Taylor's formula to the integrand, we obtain that 
\begin{align*}
    &\mathcal{U}(X+\epsilon\,\xi,Y+\epsilon\,\eta)=\;\mathcal{U}(X,Y)+\epsilon\, \delta\,\mathcal{U}+\frac{\epsilon^2}{2}\,\delta^2\,\mathcal{U}+\textit{O}(\epsilon^3)\\
    &\delta\,\mathcal{U}=\int_0^{S_t}\Big(\xi G_X+\xi' G_{X'}+\eta G_Y+\eta' G_{Y'}\Big)dS.
\end{align*}
We call $\epsilon\delta\,\mathcal{U}$ the first variation of the energy functional $\mathcal{U}$. For $X(S),Y(S)$ to be a critical point of $\mathcal{U}$, we infer that $\delta\,\mathcal{U}=0$, i.e.,
\begin{align}
    \int_0^{S_t}\Big(\xi G_X+\xi' G_{X'}+\eta G_Y+\eta' G_{Y'}\Big)dS=0,
    \label{eq:firstvariation}
\end{align}
for otherwise, we could increase or decrease the value of $\mathcal{U}$ by choosing $\epsilon$ to be of the same or a different sign of the integral in~\eqref{eq:firstvariation}, respectively.

Using integration by parts and assuming continuous derivatives of the functions involved, we arrive at the Euler-Lagrange equations:
\begin{align}
    G_X+\frac{dG_{X'}}{dS}=0,\qquad  G_Y+\frac{dG_{Y'}}{dS}=0
    \label{eq:ELequation}
\end{align}

Due to the homogeneity condition~\eqref{eq:homogeneouscondition}, the two equations in~\eqref{eq:ELequation} are not independent of each other, as we proceed to show. Differentiating~\eqref{eq:homogeneouscondition} with respect to $k$, and substituting $k=1$, yields
\begin{equation}
X'G_{X'} + Y'G_{Y'} = G.
\label{eq:connect}
\end{equation}
Differentiating this expression with respect to $X'$ or $Y'$, we obtain
\begin{equation}
\frac{1}{Y'^{2}}G_{X'X'} = -\frac{1}{X'Y'}G_{X'Y'} = \frac{1}{X'^{2}}G_{Y'Y'} = G_{1},
\label{eq:F1}
\end{equation}
where $G_{1}$ is the common value of these expressions. Differentiating~\eqref{eq:connect} partially with respect to $X$ and $Y$ we get
\begin{equation}
G_{X} = X' G_{XX'} + Y' G_{XY'},\qquad G_Y=X'G_{X'Y}+Y'G_{Y'Y} .
\label{eq:partialXderivation}
\end{equation}
Using~\eqref{eq:F1},\eqref{eq:partialXderivation} in the Euler-Lagrange equation and assuming that $X'$ and $Y'$ don't vanish simultaneously in the interval $[0,S_{t}]$ we arrive at
\begin{equation}
G_{XY'} - G_{YX'} - G_{1}(Y'X'' - X'Y'') = 0.
\label{eq:WeierstrassFirstVariation}
\end{equation}

This equation is the \textit{Weierstrass' form of the Euler-Lagrange equation}. Evaluating the necessary terms we have that
\begin{align}
G_{1} = \frac{1}{\left( X'^2 + Y'^2 \right)^{3/2}}, \label{eq:g1condition}\\
G_{XY'} = Y - H ,\qquad G_{YX'} = 0.
\end{align}
So the Euler-Lagrange equation becomes:
\begin{equation}
H - Y= \frac{X'Y'' - X''Y'}{\left( X'^{2} + Y'^{2} \right)^{3/2}}.
\label{eq:YoungLaplace}
\end{equation}

\subsection{Analytic form and family of solutions}

The differential equation~\eqref{eq:YoungLaplace} together with an initial condition determines the critical curve, but not the functions $X(S)$ and $Y(S)$. In order to find these functions we must add a second equation or differential relation between $S,X,Y$. This additional relation should be such that $X$ and $Y$ come out as single valued functions of $S$. In order to find analytic solutions to the parametric Young-Laplace equation, we make the transformation
\begin{equation}
    X'(S)=\cos{\Omega(S)},\qquad Y'(S)=\sin{\Omega(S)},
    \label{eq:ode12}
\end{equation}
where $\Omega$ is the tangential angle to the meniscus. These substitutions define the independent variable $S$ to be the arc length of the meniscus and turn the Young-Laplace equation into 
\begin{equation}
    \Omega'(S)=H-Y.
    \label{eq:ode3}
\end{equation}
This transformation splits the Young-Laplace equation into a system of 3 ODEs. We set the initial conditions of the meniscus to have general pinning conditions
\begin{equation*}
    X(0)=0,\qquad Y(0)=0,\qquad \Omega(0)=\theta.
\end{equation*}
To find an analytic solution to the system of ODE's we differentiate \eqref{eq:ode3} and substitute \eqref{eq:ode12} to find
\begin{equation}
    \Omega''(S)=-\sin{\big(\Omega(S)\big)}.
    \label{eq:ode4}
\end{equation}
We observe that the dynamics of the tangent angle ($\Omega$) are similar to the dynamics of a pendulum or an elastic rod~\citep[p. 265]{levien2009spiral}. Multiplying \eqref{eq:ode4} with $\Omega'(S)$ and integrating, we get,
\begin{equation}
    \frac{1}{2}\Omega'^2-\cos{\Omega}=A,
    \label{eq:energy_constant}
\end{equation}
where $A$ is the constant of integration. For a simple pendulum, $A$ defines the energy of the system. In our case, $A$ represents the horizontal force balance of hydrostatic pressure, $\frac{1}{2}\rho g (h-y)^2$, and surface tension, $\sigma \cos \Omega$, along any longitudinal cross section of the meniscus. Using the initial conditions $\Omega(0)=\theta$ and $\Omega'(0)=H$, we evaluate the integration constant as,
\begin{equation}
    A=\frac{H^2}{2}-\cos{\theta}.
    \label{eq:constantA}
\end{equation}
Using the trigonometric identity $\cos{\Omega}=1-2\sin^2\Omega/2$, we arrive at
\begin{equation}
    \Omega'(S)=2\sqrt{\frac{A+1}{2}-\sin^2\frac{\Omega}{2}}.
    \label{eq:omega_dash}
\end{equation}
The solution to this differential equation can be explicitly written down in terms of Legendre elliptic and Jacobi amplitude functions,
\begin{equation}
    \Omega(S)=2\;\text{am}\Big(\sqrt{\frac{1+A}{2}}S+\mathit{F}\Big(\frac{\theta}{2}\Big|\frac{2}{1+A}\Big)\Big|\frac{2}{1+A}\Big),
\end{equation}
where $\mathit{F}(u|m)$ and is the incomplete elliptic integrals of the first kind and $\text{am}(u|m)$ is the Jacobi amplitude function. $Y(S)$ can be calculated directly using \eqref{eq:ode3} and the identity $\text{am}(u,k)=\int_0^u\text{dn}(u',k)du'$ to be
\begin{equation}
    Y(S)=H-\sqrt{2(1+A)}\;\text{dn}\Big(\sqrt{\frac{1+A}{2}}S+\mathit{F}\Big(\frac{\theta}{2}\Big|\frac{2}{1+A}\Big)\Big|\frac{2}{1+A}\Big),
    \label{eq:Y}
\end{equation}
where $\text{dn}(u|m)$ is the Jacobi delta amplitude function. Using the result \eqref{eq:ode3}, \eqref{eq:energy_constant} and \eqref{eq:Y} into the expression for $X'(S)$ in \eqref{eq:ode12} we obtain
\begin{align}
    X(S)&=\sqrt{2(1+A)}\;\mathit{E}\Big(\text{am}\Big(\sqrt{\frac{1+A}{2}}S+\mathit{F}\Big(\frac{\theta}{2}\Big|\frac{2}{1+A}\Big)\Big|\frac{2}{1+A}\Big)\Big)\Big|\frac{2}{1+A}\Big)\nonumber\\
    &-\sqrt{2(1+A)}\;\mathit{E}\Big(\frac{\theta}{2}\Big|\frac{2}{1+A}\Big)-AS.
    \label{eq:X}
\end{align}
$\mathit{E}(u|m)$ is the incomplete elliptic integral of the second kind. 

\begin{figure}
  \centerline{\includegraphics[width=.9\textwidth]{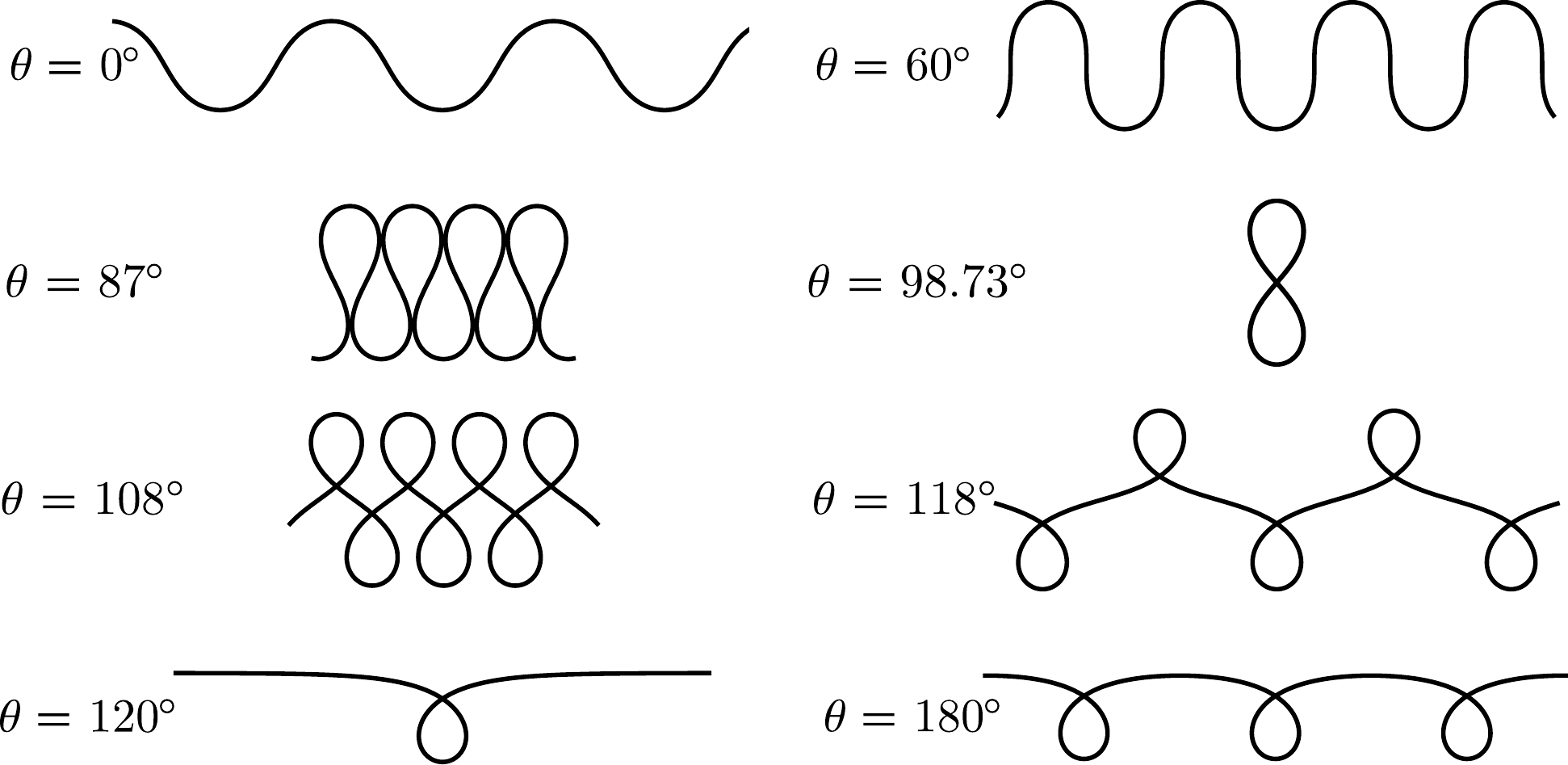}}
  \caption{Family of solutions for the parametric Young-Laplace equation}
  \label{fig:family_of_solutions}
\end{figure}

Using these analytic expressions, we plot the interface for different values of pinning angle $\theta$ as shown in figure~\ref{fig:family_of_solutions}. For the purpose of this illustration, we consider the non-dimensional melt height $H$ to be 1 as the pinning angle $\theta$ is varied. Figure~\ref{fig:family_of_solutions} shows parametric plots of $X(S)$ and $Y(S)$ at $H=1$ and for some chosen values of $\theta$ using \eqref{eq:Y} and \eqref{eq:X}. The curves are periodic and can extend in length from $(-\infty,\infty)$. The origin can therefore be taken to be any point where the curve makes an angle $\theta$ with the horizontal. Plotting for $A>1$ might require certain inversion transformations. These along with other identities in this section can be found in \citet{abramowitz1948handbook}.  

Certain sections of the curves in Figure~\ref{fig:family_of_solutions} closely resemble pendant or droplet like shapes. The similarity of these solutions also extend to the family of elastic curves discovered by Euler as a part of his elastica problem~\citep{maxwell1876capillary}. Depending on the second order stability condition, a portion of these curves will form a stable meniscus to the ribbon growth process. This condition will be derived in the following section.

\section{Stability analysis via the second variation}
In this section we consider the stability of the critical curves when the end-points are considered fixed. Using Taylor series representation, the second variation in parametric form is expressed as:
\begin{equation}
\delta^{2}U_{0} = \int_0^{S_{t}} \delta^{2} G\, dS,
\end{equation}
where
\begin{equation}
\begin{split}
\delta^{2}G = G_{XX}\xi^2 + 2G_{XY}\xi\eta + G_{YY}\eta^{2} + 2G_{XX'}\xi\xi' + 2G_{YY'}\eta\eta' +\\
2G_{XY'}\xi\eta' + 2G_{YX'}\eta\xi' + G_{X'X'}\xi'^{2} + 2G_{X'Y'}\xi'\eta' + G_{Y'Y'}\eta'^{2}.
\label{eq:integrand}
\end{split}
\end{equation}
In order for the curve described by $Y(S)$ and $X(S)$ to be a minimum---and therefore stable---its second variation should be positive; so the value of the integral above must be always positive in the range of integration. Using a lengthy factorization, Weierstrass transformed the second variation into the classical quadratic functional
\begin{equation}\label{eq:delU}
\delta^{2}U_{0} = \int_0^{S_{t}} \left[G_{1}\left(\frac{d\omega}{dS}\right)^{2} + G_{2}\omega^{2} \right]\, dS.
\end{equation}
In our application we find from \eqref{eq:integrand}
\begin{equation*}
\omega = Y'\xi - X'\eta,
\end{equation*}
and $G_{2}$ satisfies the following relationships:
\begin{equation*}
G_{2} = \frac{L_{2}}{Y'^{2}} = \frac{M_{1}}{-X'Y'} = \frac{N_{1}}{X'^{2}},
\end{equation*}
with
\begin{eqnarray*}
L_{2} &=& G_{XX} - Y''G_{1} - \frac{dL_{1}}{dS}, \\
M_{2} &=& G_{XY} + X''Y''G_{1} - \frac{dM_{1}}{dS},\\
N_{2} &=& G_{YY} - X''^{2}G_{1} - \frac{dN_{1}}{dS}, \\
L_{1} &=& G_{XX'} - Y'Y''G_{1}, \\
M_{1} &=& G_{XY'} + X'Y''G_{1} = G_{YX'} + Y'X''G_{1}, \\
N_{1} &=& G_{YY'} - X'X''G_{1}.
\end{eqnarray*}
The form of the integral \eqref{eq:delU} allowed Weierstrass to apply the classical results of the calculus of variations. Namely, Legendre's necessary condition and Jacobi's test. Legendre's necessary condition for a minimum requires that 
\begin{equation}
G_{1} \geq 0,
\end{equation}
along the stationary curve described by $X(S)$ and $Y(S)$. 

Jacobi's test requires that the solution to the differential equation,
\begin{equation}
G_{2} u - \frac{d}{dS}\left(G_{1}\frac{du}{dS} \right) = 0,
\end{equation}
must not have conjugate points in the integration interval, i.e:
\begin{equation*}
u(S) \neq  0 \hspace{5 mm} \hspace{5mm} 0 < S < S_{t}.
\end{equation*}
For otherwise if $u(S)=0$, it is possible to find a perturbation centered around $S$ such that the second variation is negative. 

The Legendre test is satisfied by virtue of \eqref{eq:g1condition}. In order to have the Jacobi test satisfied we require that the solution to the differential equation
\begin{equation}
G_{2}u - (G_{1}u')' = \frac{Y'''}{Y'}u - u''= 0,
\label{eq:JacobiWeierstrass}
\end{equation}
must not have conjugate points in the interval of integration. Equation~\eqref{eq:JacobiWeierstrass} is equivalent to:
\begin{equation*}
(Y'u' - Y''u)' = 0.
\end{equation*}
Thus, we obtain that for some constant $K_1$, 
\begin{equation*}
Y'u' - Y''u = K_{1}.
\end{equation*}
Dividing the expression above by $Y'^{2}$ we get:
\begin{equation*}
\frac{Y'u' - Y''u}{Y'^{2}} = \left( \frac{u}{Y'} \right)' = \frac{K_{1}}{Y'^{2}}.
\end{equation*}
So the condition for stability becomes 
\begin{equation}
u(S) = K_{1}Y'(S)\int_0^{S}\frac{dS}{Y'(S)^{2}} \neq 0 \text{    for    } 0 < S < S_{t}.
\label{eq:StabilityCriterionWeierstrass}
\end{equation}
The integral in \eqref{eq:StabilityCriterionWeierstrass} is always positive as long as $Y'(S) \neq 0$ in the interval $0<S<S_t$ (given $S_{t} > 0$), otherwise the integral does not converge. The term $K_{1}Y'(S)$ does not change sign as long as $Y'(S)$ does not change sign in the $(0,S_{t})$ interval. Therefore the issue of stability reduces to finding the range of values for which $Y'(S)$ crosses zero in the range $(0,S_{t})$. This is easier to analyse recalling from \eqref{eq:ode12} the fact that $Y'(S) = \sin\Omega(S)$, where $\Omega(S)$ is the tangential angle of the meniscus with respect to the horizontal axis. If $\sin\Omega(S)$ is always positive or always negative in the integration interval, the function $u(S)$ will have no conjugate points and the Jacobi test is satisfied. Thus we simplify our stability criterion \eqref{eq:StabilityCriterionWeierstrass} to the following expression:
\begin{equation}
\sin(\Omega(S)) \neq 0 \qquad \forall \text{   }S  \in (0,S_{t}),
\end{equation}

\section{Results} \label{results}

We apply the theory developed in the previous sections to study the properties of menisci in a silicon ribbon growth process while keeping in mind that they can be applied to a range of other problems of physical and engineering importance such as finding the size and stability properties of droplets. To characterize figure~\ref{fig:HRG_meniscus} in more detail, the edge of the crucible is considered to be rectangular ($\phi=90^\circ$). One end of the meniscus is considered to remain pinned at the edge of the crucible and the other end to intersect the ribbon at a fixed angle ($\sigma$). The nature of the angle ($\sigma$) depends on whether solidification or melting takes place at the triple point, in which case $\sigma$ is either the growth angle or the contact angle respectively. To maintain consistency in our analysis and for the sake of convenience, we assume $\sigma$ to be a constant value of ($11^\circ$) across all operating conditions. The pull angle ($\beta$) and the height of the melt ($h$) serve as degrees of freedom. It is of interest to find stable operating regimes for the meniscus over the parameter space of $\beta$ and $h$. The required material properties are summarized in table~\ref{tab:properties} and are used to dimensionalize the equations and the results.

\begin{figure}
  \centerline{\includegraphics[width=\linewidth]{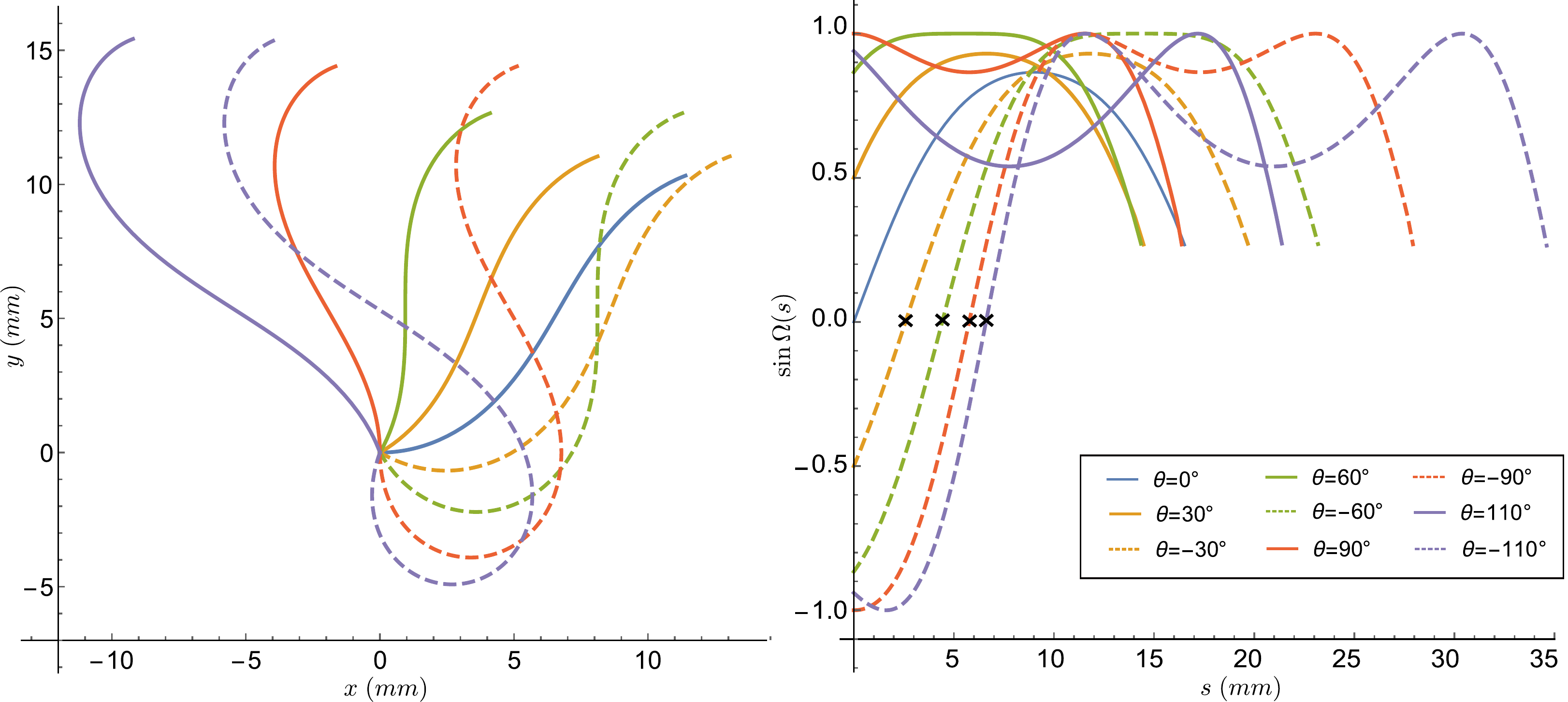}}
  \caption{(left) Stationary meniscus shapes obtained using the analytical solution in parametric form. Curves correspond to a value of $\beta=10^\circ$ and $h=5.35\,mm$. (right) The sine of the tangent angle $\Omega(s)$ for different pinning angles. The curves crossing zero correspond to the unstable modes.}
  \label{fig:many_curves}
\end{figure}

\begin{table}
\begin{center}
\begin{tabular}{ccc}
\toprule
Parameter	&Symbol   &Value\\[3pt]
\midrule
Density of liquid silicon		&$\rho$		&2570 $[\text{kg} \hspace{1mm} \text{m}^{-3}]$\\
Acceleration due to gravity			&$g_r$			&9.8  $[\text{m} \hspace{1mm} \text{s}^{-2}]$\\
Surface tension of silicon		&$\gamma$			&0.72 $[\text{J} \hspace{1mm} \text{m}^{-2}]$\\
Triple point angle	&$\sigma$							&$11^{\circ}$\\
Melt-graphite wetting angle	 &$\theta_{e}$		&$30^{\circ}$\\
\bottomrule
\end{tabular}
\caption{Material properties and parameters used in the illustrative example.}
\label{tab:properties}
\end{center}
\end{table}

The plot on the left in Figure~\ref{fig:many_curves} describes the various stationary meniscus shapes for a representative pulling angle of $\beta=5^\circ$ and melt height $h=5.35\,\text{mm}$ ($H=1$). We use the analytic expressions for $x(s)$ and $y(s)$, with different $\theta$ values to plot the interface curves and stop when the interface reaches the angle of $\sigma+\beta$. In order to show the concept of static stability, we focus on the results obtained from Jacobi's test (Legendre's condition for a minimum is always satisfied for all meniscus shapes). The plot on the right in Figure \ref{fig:many_curves} shows the sine of the tangential angle as a function of the arc length. As we mentioned before, the sine of the tangential angle must not vanish between 0 and $s_t$. From the figures we show that menisci in which the pinning angles are greater than zero are statically stable, whereas the curves for values of pinning angle lower than zero cross the horizontal axis. The family of stable and unstable curves converge in the limit $\theta\rightarrow 0$.

Let $x^*$ and $y^*$ be the parametric coordinates describing the equation for a ribbon. We assume the shape of the ribbon to be a straight line starting from $l_c=-5.35\text{cm}$ $(L_c=-10)$ and represented by
\begin{equation*}
    \mathit{L}(x^*,y^*)=(y^*-h)-\tan{\beta}(x^*-l_c)=0.
\end{equation*}
The desired solution is then given by any curve described in Figure \ref{fig:many_curves}, whose end point lies on this line. This can be formulated as a boundary value problem:
\begin{align}
    &x(0)=0\qquad\qquad\; y(0)=0\nonumber\\
    &\Omega(s_t)=\beta+\sigma\qquad L(x(s_t),y(s_t))=0
    \label{eq:bvp}
\end{align}

\begin{figure}
  \centerline{ \includegraphics[width=\linewidth]{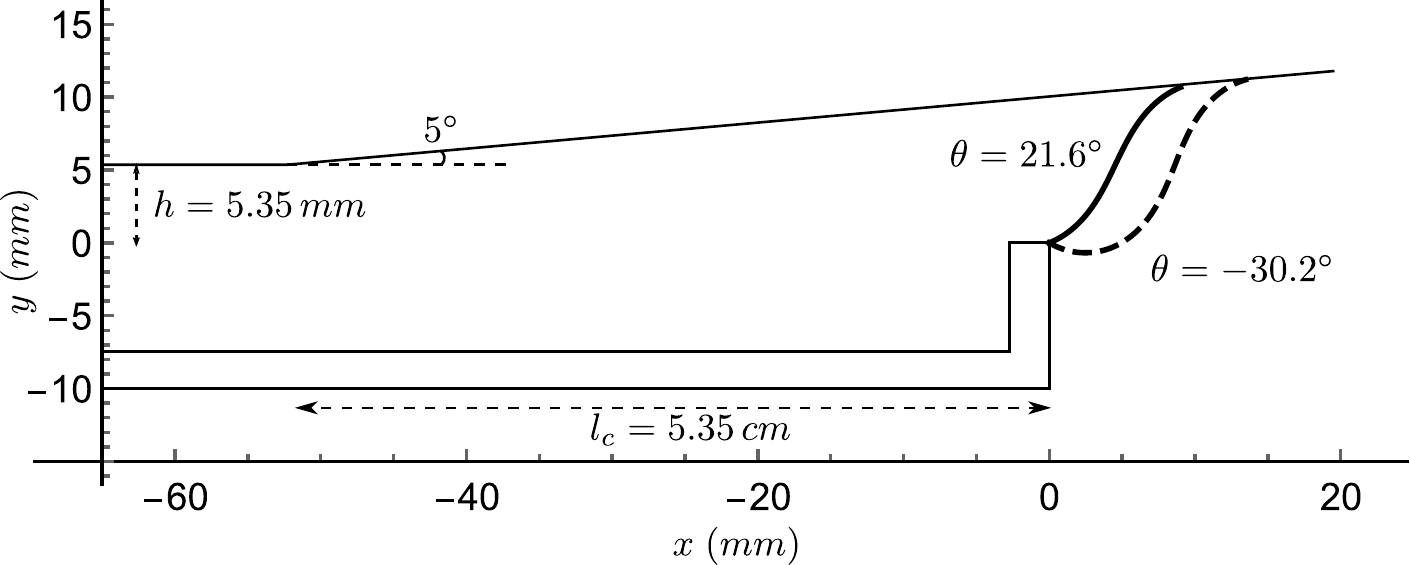}}
  \caption{Hydro-statically stationary configuration for a melt level of $5.35\,mm$ and a pulling angle of $5^\circ$. The solid curve corresponds to a statically stable configuration and the dashed curve corresponds to an unstable configuration.}
  \label{fig:to_scale}
\end{figure}

 We use a Newton-Raphson solver to find curves that satisfy \eqref{eq:bvp}. Two curves, one stable and one unstable, are found and illustrated in Figure~\ref{fig:to_scale} along with a diagram of the system (to scale) to better visualize the concept of hydro-static stability.

\subsection{Effect of pull angle} \label{b_effect}

\begin{figure}
  \centerline{\includegraphics[width=\linewidth]{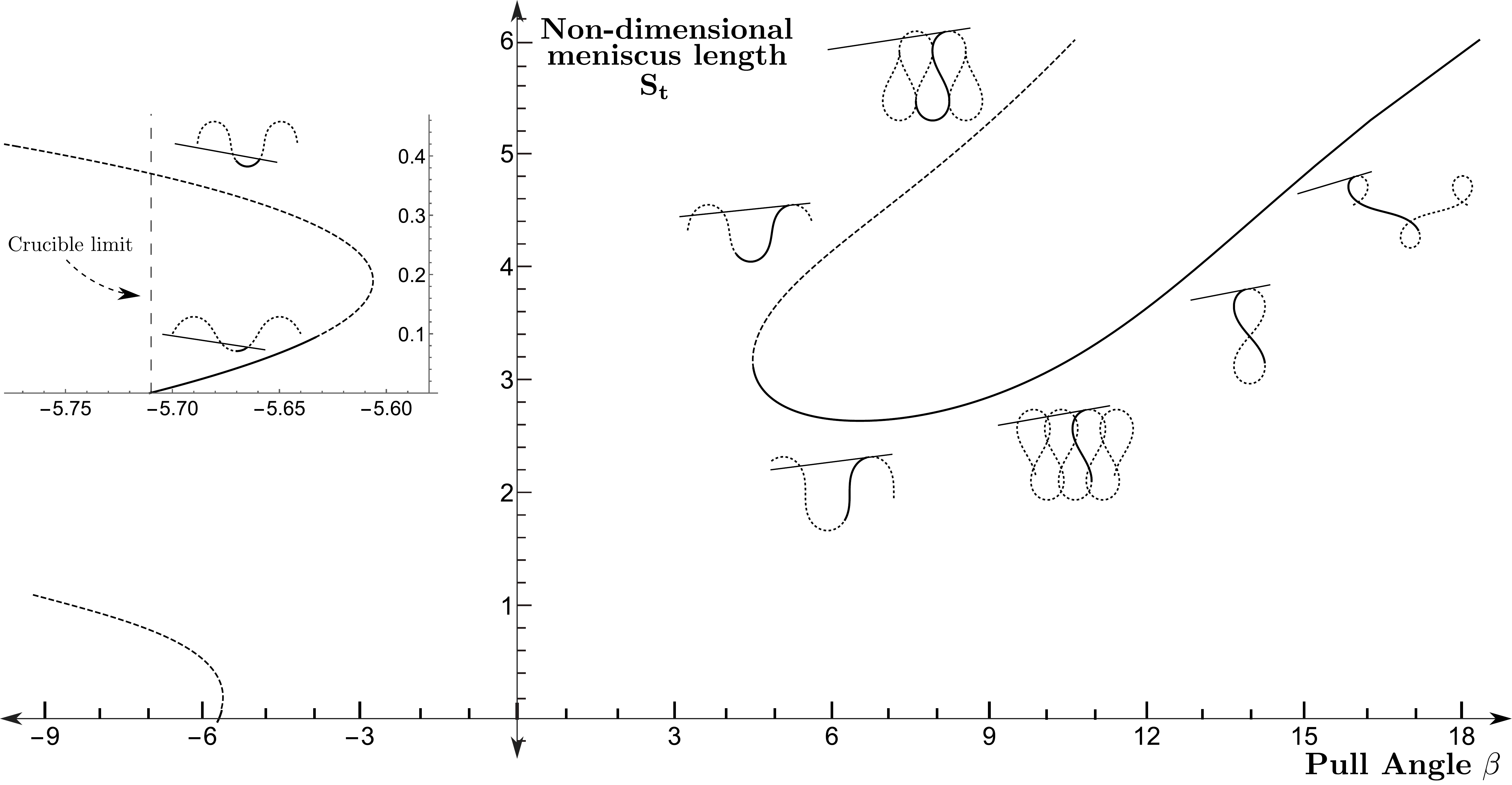}}
  \caption{Saddle node bifurcations in the meniscus length $S_t$ and pull angle $\beta$ solution space. Inset: A zoomed up diagram of the solution space for negative pull angles. Solid lines represent stable solutions for which the Jacobi test is satisfied.}
  \label{fig:bifurcation}
\end{figure}

The shape of the meniscus is influenced by the pull angle ($\beta$) through the boundary conditions described in \eqref{eq:bvp}. To get a better description of the meniscus multiplicity observed above, we vary the pulling angle to evaluate the feasibility of stationary menisci. Figure~\ref{fig:bifurcation} provides a description of the solution space for a melt height of $h=5.35\,mm$ as a function of the pulling angle. The choice of meniscus length as the Y-axis was motivated from literature on elasticity and bifurcation theory~\citep{marsden1994mathematical}. 

Representative meniscus shapes are drawn along the solution curves to describe their geometry for a few choice of pull angles. The dashed curves describe the family of unstable solutions, characterized by a point of zero slope where the Jacobi condition is not met. In the neighbourhood of this point, it is possible to perturb the curve such that the second order variation is negative and the solution is not a minimum. Vice versa, the solid curves describe the statically stable solutions which minimize the thermodynamic energy of the system. The pinning condition at the crucible edges due to Gibbs has not been considered here and is commented on separately in Appendix~\ref{appendix}.

We observe that it is not always possible to find a feasible solution for any given value of pull angle. Such an operational limit was also observed in the thermal-capillary simulations performed by \citet{Daggolu2013}, however their analysis was limited to the narrow stability region on the left. Two saddle node bifurcations are observed in our analysis that divide the feasible solution space into two disjoint regions. The feasible region on the left has a smaller range of pull angles available for stable operation. The crucible limit shown inset is the limit at which the meniscus length goes to zero. Decreasing the pulling angle to this limit causes the bottom part of the ribbon to get closer to the crucible edge and result in ribbon freezing onto the crucible. On the other hand, increasing the pull angle beyond the bifurcation point results in the meniscus becoming unstable and causing the melt to spill-over from the crucible. 

Given the narrow range of operation for negative pull angles, it would be desirable to operate the ribbon growth process at positive pull angles, beyond $4.5^\circ$ for the case of $H=1$, as there is no upper limit to the height of the pulling angles. The feasible region on the right illustrates the variety of meniscus shapes that can be achieved for positive pull angles. The low-angle silicon sheet (LASS) growth process, where the ribbon is extracted from the melt at a slight positive angle with the horizontal takes advantage of this idea.~\citep{jewett1982progress}. 

\subsection{Effect of melt height} \label{h_effect}
 
A successful design for a ribbon growth process requires understanding the effect of the melt height ($H$) on the stability of the meniscus. It is therefore useful to study how the melt height influences the landscape of the solution space described in Section~\ref{b_effect}. 

\begin{figure}
  \centerline{\includegraphics[width=\linewidth]{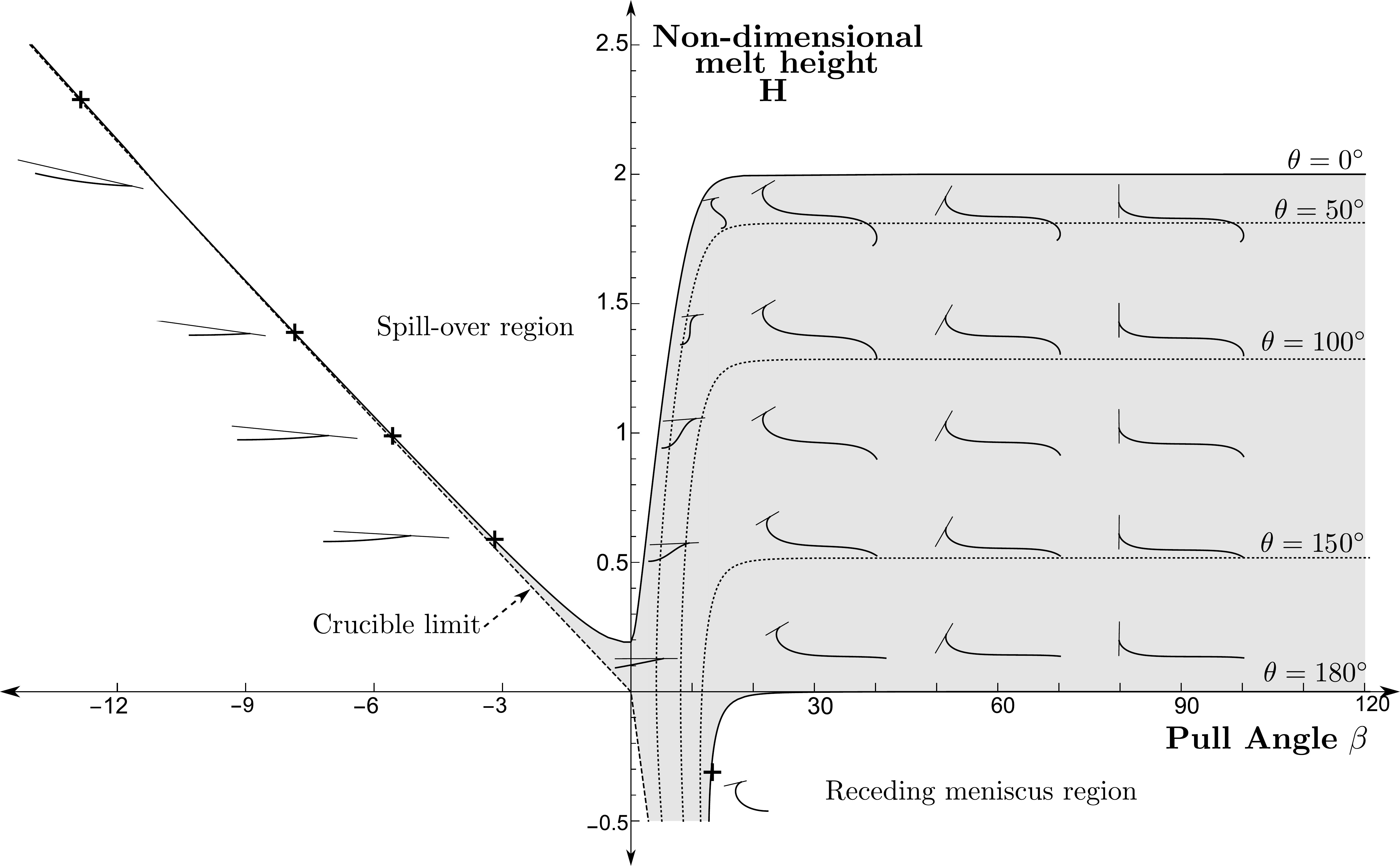}}
  \caption{Stable meniscus region (shaded) over the parameter space of melt height and pull angle. Representative meniscus shapes have been drawn for melt height and pull angles.}
  \label{fig:hvb}
\end{figure}

Figure~\ref{fig:hvb} shows the range of feasible pull angles as the melt height is varied. For example, the portion of the shaded region at $H=1$ can be thought of as a projection of the feasible pull angles (solid curves) in Figure~\ref{fig:bifurcation} onto the X-axis. Therefore, the entire shaded region in Figure~\ref{fig:hvb} describes the existence of a stable meniscus at every point over the parameter space of $H$ and $\beta$.
 
Representative meniscus shapes have been drawn for some chosen values of $H$ and $\beta$. At some places, a plus symbol has been used to denote the point where the menisci belong. For $H<2$ ($h<10.7\,mm$) we see that it is possible to find a stable meniscus for pull angles as large as $90^\circ$. At this point the arrangement corresponds to vertical ribbon growth techniques like WEB, EFG. What is interesting to note is that as the pull angle increases, the meniscus becomes longer and the meniscus-ribbon triple point moves further away from the crucible edge. This observation is the guiding principle behind low-angle silicon sheet (LASS) growth process and circumvents the problem of ribbon freeze-over by moving the triple phase contact point on the ribbon away from the crucible edge.
 
As the melt height increases, we see that above $H=0.2$, the feasible solution space splits. The portion in between the regions is the melt spill-over region. In this region it is not possible to form a stable meniscus to support the melt from spilling over the crucible. Since the solution space for positive pull angles is much larger than the negative pull angles, the scale for the negative pull angles has been increased to meaningfully show the feasible solution space. Notice also that there is an upper limit to the height of the melt that the meniscus can accommodate. Beyond this height, a meniscus can no longer exist and the melt spills over from the crucible edge.

The feasibility region shown in Figure~\ref{fig:hvb} does not consider the Gibbs inequality condition that arises at the crucible edge. Gibbs' inequality provides a range of pinning angles ($\theta$) at the crucible edge for which the meniscus remains stable. Since the Gibbs limit is a material property and also depends on the geometry of the crucible edge, we provide contours for some chosen pinning angles to find a subset of the feasible (shaded) region that satisfies Gibbs' inequality.   

\section{Experimental Design}

A miniature proof-of-concept experiment is used to study and illustrate the mechanism of melt spill over in a HRG configuration when the pull angle is varied. A polyethylene ribbon ($\rho_{pe} = 0.93g/cm^3 < \rho_{water}$) rests completely on top of the water contained in a plastic bath such that the inclination with respect to the top surface of the water can be varied. 

\begin{figure}
  \centerline{\includegraphics[width=\linewidth]{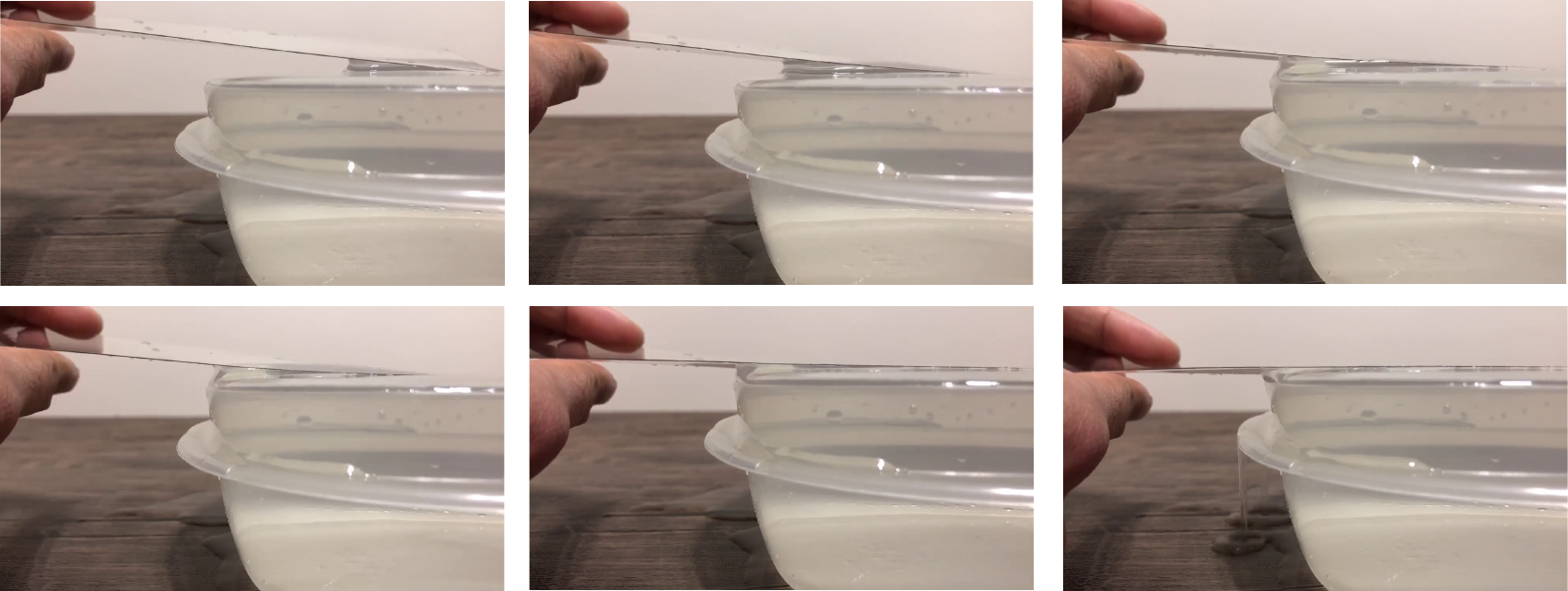}}
  \caption{A sequence of photographs showing the meniscus spilling over from the corner of the plastic bath. Spill-over is induced by decreasing the angle of the sheet with the bath}
  \label{fig:experiment}
\end{figure}

From Section~\ref{h_effect}, we observed a range of infeasible pull angles around the horizontal position ($\beta=0$) when $H$ was greater than $0.2$. To test this hypothesis, we induce spill-over by slowly decreasing the angle of inclination with the water surface while photographing the changes in the shape of the water meniscus. Figure~\ref{fig:experiment} displays a sequence of photographs showing the bulking of the meniscus as the pulling angle decreases. The top left-most photograph shows the shape of a meniscus in which the ribbon is inclined at a positive angle (a stable configuration). This configuration would make it least likely for the ribbon to freeze on top of the crucible edge. The right-most bottom photograph is the shape of a meniscus prior to spilling over the crucible (an unstable configuration) as the ribbon becomes horizontal. Despite the difference in materials, we see that the stability analysis from our theory agree qualitatively with the experimental observations.

\section{Conclusions}

This paper provides a parametric formulation and a solution to the generalized static stability problem for the meniscus in a ribbon growth process. Due to the geometric nature of the meniscus problem, we observe that the method of parametric representation is not only preferable but also one which furnishes a complete solution. Using Weierstrass' variational theory, we found analytic expressions describing the shape of the meniscus and compared it with the family of Euler's elastic curves. This similarity can be used to exchange concepts from elasticity theory in order to study stability and bifurcations of menisci shapes in liquids and vice versa.

The stability of the meniscus is evaluated using Legendre and Jacobi test conditions. A range of stable operating conditions are provided over the parameter space of melt height and pull angle. Two bifurcation points are observed which divide the solution space into two regions. The infeasibility zone between the two solution spaces, which include the horizontal position of the ribbon, didn't have a stable meniscus solution to support the ribbon. Growing a ribbon in this region leads to melt spilling over from the meniscus until the melt height decreases to the stable region. This argument is supported by doing a simple proof of concept experiment, in which the phenomena of spill-over is created using a polyethylene ribbon resting on a bath of water. Given the vast range of stable positive pull angles, we conclude that it is appropriate to incline the ribbon above a certain threshold angle to ensure stability of ribbon growth as the horizontal configuration was statically unstable.


\section*{Acknowledgement}
This research has been funded by the National Science Foundation (No. CBET 093 2556).

\bibliographystyle{plainnat}
\bibliography{jfm-instructions}

\clearpage
\appendix
\section{Young's contact angle and Gibbs pinning condition} \label{appendix}

\begin{figure}
  \centerline{\includegraphics[width=0.5\linewidth]{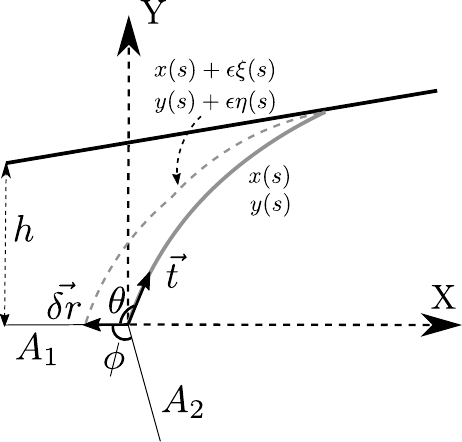}}
  \caption{For the case of variable endpoints, the perturbation at the origin is considered to be $\vec{\delta r}$. The aim is to find conditions on the meniscus shape such that $\vec{\delta r}=0$ is a minimum.}
  \label{fig:appendix_schematic}
\end{figure}

The parametric Young-Laplace equation \eqref{eq:YoungLaplace} derived in section~\ref{first_variation} relies on contact angle conditions at the boundary in order to find stationary curves that describe the meniscus shape. By perturbing the end points of the meniscus, we show that the contact angle and the Gibbs pinning conditions follow as a consequence of the free energy minimization of the system.

\begin{figure}
  \centerline{\includegraphics[width=0.7\linewidth]{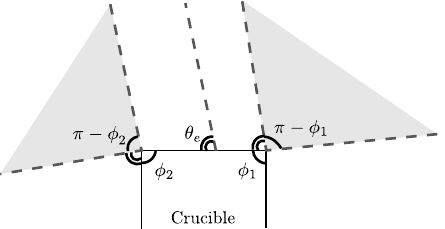}}
  \caption{The shaded region displays the range of pinning angles for the inner and the outer edges. The figure shows how the starting position of the meniscus would change as the pinning angle is varied.}
  \label{fig:appendix_cases}
\end{figure}

Consider the free energy formulation for the meniscus $x(s),\,y(s)$ as defined in section~\ref{reformulation} and add the surface energy of the solid boundaries in contact with the air and the liquid. For simplicity, we briefly consider the case where only the end point at the origin is varied while the end point at $s_t$ is considered fixed. In this case,
\begin{equation*}
    \Delta U=\int_0^{s_t}-\Delta P \:xy'ds+\gamma \sqrt{x'+y'}ds+ A_1\gamma_1+A_2\gamma_2,
\end{equation*}
where $A_1$ and $A_2$ are the areas of the solid crucible in contact with the melt and the air respectively. $\gamma_1$, $\gamma_2$ are the interfacial energies of the melt and the air boundaries with the crucible (see figure~\ref{fig:appendix_schematic}). 

As before, we introduce small perturbations of $\epsilon \xi(s)$ and $\epsilon \eta(s)$ into $x(s)$ and $y(s)$. These perturbations are fixed at $s_t$ such that $\xi(s_t)=\eta(s_t)=0$. Using the definition \eqref{eq:freeenergy}, the first variation in energy is given by
\begin{equation*}
    \delta U =   \int_0^{s_t} (\xi g_x+\xi'g_{x'}+\eta g_y+\eta' g_{y'}) ds+ \delta r(\gamma_2-\gamma_1),
\end{equation*}
where $|\delta r|=\sqrt{\xi(0)^2+\eta(0)^2}$. Integrating by parts, we arrive at the following form of the first variation,
\begin{equation*}
    \delta U =   \int_0^{s_t}\Bigg[ \xi \Big(g_x-\frac{d}{ds}g_{x'}\Big)+\eta \Big(g_y-\frac{ d}{ds}g_{y'}\Big)\Bigg] ds- \xi g_{x'}\Big|_0-\eta g_{y'}\Big|_0+ \delta r(\gamma_2-\gamma_1).
\end{equation*}

Setting the integrand to zero gives us the Euler-Lagrange equation for optimality. Substituting the expressions for $g_{x'}$ and $g_{y'}$ give us
\begin{equation*}
    \delta U=-\gamma \frac{\xi x'+ \eta y'}{\sqrt{x'^2+y'^2}}\Bigg|_0+\delta r (\gamma_2-\gamma_1).
\end{equation*}

The first term can be interpreted as a dot product between $ \vec{\delta r}=[\xi(0),\eta(0)]$ and $\vec{t}=[x'(0),y'(0)]$, which can be written in terms of the cosine of the angle between them.
\begin{equation*}
    \delta U=
    \begin{cases}\gamma \delta r \Big(-\cos(\theta) +\frac{\gamma_2-\gamma_1}{\gamma}\Big) & \delta r>0\\
    \gamma \delta r \Big(\cos(2\pi-\phi-\theta) +\frac{\gamma_2-\gamma_1}{\gamma}\Big) & \delta r<0
    \end{cases}
\end{equation*}
We see that the first variation is minimized and becomes zero at $\delta r=0$ when
\begin{align*}
\theta_e\leq\theta\leq\pi-\phi+\theta_e\,,\\ \theta_e=\arccos\Big(\frac{\gamma_2-\gamma_1}{\gamma}\Big).
\end{align*}
This range of $\theta$ values is known as the Gibbs pinning condition and is illustrated by the shaded region in figure~\ref{fig:appendix_cases}. The Young-Dupre contact angle condition follows as a corollary by setting $\phi=\pi$.  

When $\theta\leq\theta_e$ the meniscus recedes horizontally along the crucible boundary until it gets pinned to the inner corner of the crucible. An extended range of pinning angles, as illustrated by the shaded region in figure~\ref{fig:appendix_cases} follows. The overall range for the pinning conditions can be derived using a similar analysis.

\end{document}